\documentclass[aps,prb,twocolumn,superscriptaddress, showspace,showkeys,floatfix]{revtex4}
\usepackage{amsmath}
\usepackage{bm}
\usepackage{amssymb}
\usepackage{graphicx}
\usepackage{epstopdf}
\usepackage{bmpsize}
\usepackage{epsfig}
\usepackage{color}
\usepackage{hyperref}
\usepackage{float}
\usepackage{braket}
\usepackage{lipsum}

\begin{document}
\title{Cavity QED Based on Thermal Atoms Interacting with a Photonic Crystal Cavity: A Feasibility Study}

\date{\today}

\author{Hadiseh Alaeian}

\author{Ralf Ritter}

\author{Muamera Basic}

\author{Robert L\"ow}

\author{Tilman Pfau}
\affiliation{5 Physikalisches Institut, University of Stuttgart, Pfaffenwaldring 57, 70569 Stuttgart, Germany} 

\affiliation{Center for Integrated Quantum Science and Technology (IQST) in Stuttgart and Ulm, Germany}

\begin{abstract}
The paradigm of cavity QED is a two-level emitter interacting with a high quality factor single mode optical resonator. The hybridization of the emitter and photon wave functions mandates large vacuum Rabi frequencies and long coherence times; features that so far have been successfully realized with trapped cold atoms and ions and localized solid state quantum emitters such as superconducting circuits, quantum dots, and color centers~\cite{Reiserer2015,Faraon2010}. Thermal atoms on the other hand, provide us with a dense emitter ensemble and in comparison to the cold systems are more compatible with integration, hence enabling large-scale quantum systems. However, their thermal motion and large transit time broadening is a major challenge that has to be circumvented. A promising remedy could benefit from the highly controllable and tunable electromagnetic fields of a nano-photonic cavity with strong local electric-field enhancements. Utilizing this feature, here we calculate the interaction between fast moving, thermal atoms and a nano-beam photonic crystal cavity (PCC) with large quality factor and small mode volume. Through fully quantum mechanical calculations, including Casimir-Polder potential (i.e. the effect of the surface on radiation properties of an atom) we show, when designed properly, the achievable coupling between the flying atom and the cavity photon would be strong enough to lead to Rabi flopping in spite of short interaction times. In addition, the time-resolved detection of different trajectories can be used to identify single and multiple atom counts. This probabilistic approach will find applications in cavity QED studies in dense atomic media and paves the way towards realizing coherent quantum control schemes in large-scale macroscopic systems aimed at out of the lab quantum devices.

\end{abstract}


\maketitle

\section{Introduction}
The field of cavity quantum electrodynamics (CQED) dates back to more than 50 years ago when Purcell in his seminal work reported that the radiation properties of an atom can be modified via its surroundings~\cite{Purcell1946}. Within the last decades CQED has been a versatile and powerful testbed to investigate fundamental postulates of quantum mechanics such as superposition and entanglement~\cite{Schleich1996, Raimond2001}. In addition, it has been the source of various developments in the fields of quantum technologies and quantum information~\cite{Walther2006,Haroche2007,Kimble2008,Reiserer2015}. 

Early CQED experiments were considering the modification of the atom lifetime and its radiation properties in the vicinity of a low quality factor cavity. However, with the development of high finesse cavities most of CQED studies shifted towards exploring \emph{strong coupling} regime where the energy between the atom and cavity photon is exchanged coherently. Within that regime, starting with microwave CQED, entanglement between highly excited Rydberg atoms flying accross superconducting cavities and microwave photons was observed~\cite{Davidovich1994}. Later, by combining low-lying atomic transitions and high-finesse dielectric cavities strong atom-photon couplings in the visible range and at room temparture was demonstrated ~\cite{Rempe1991,Thompson1992,Munstermann1999, Hood2000, Lee2014}. 

The introduction of nano-photonics to the field of quantum optics and atomic physics has substantially broadened the capabilities of atom-photon systems. Nano-photonic and plasmonic structures provide quantum emitters with a large tailorability of the local density of the optical states (LDOS), hence making a unique platform for studying well-designed and controllable atom-light interaction. In addition, quantum optical platforms can exploit the typical high quality factor (Q) and the small mode volume of photonic modes of nano-photonic devices to explore the very large cooperative coupling regimes. 
Such unique capabilities have been actively investigated
in various solid-state systems such as quantum dots~\cite{Englund2007, Faraon2010, Faraon2011}, color centers~\cite{Radulaski2017, Zhang2018}, and embedded rare-earth ions~\cite{Chen2016,Miyazono2017}. 
Each of these platforms has its strength and weaknesses and might be suitable only for specific problems.

However, unless particular treatments are considered, in most of these systems the coupling of quantum emitters to phonons of the host material causes large inhomogenous broadening for the emitters  which is a major bottleneck in these atom-like cavity systems. 

On the other hand, atoms are naturally identical quantum emitters so the system composed of these quantum emitters combined with nano-photonic devices would substantially improve the inhomogeneous broadening of these hybrid systems~\cite{Kien2017,Schneeweiss2017}. Therefore, the hybrid quantum systems of atoms and nano-photonic devices have a promising perspective for exploring new realms of CQED. Within the last two decades and with the advancements of nano-photonics and nano-technology, quantum optics has witnessed a lot of efforts focused on interfacing these structures with neutral atoms~\cite{Alton2011, Asenjo2017, Burgers19, Douglas2015, Goban2015, Hood2016}.

Among the various nano-photonic devices photonic crystal (PhC) cavities are some of the most promising candidates to obtain high-Q resonances and small mode volume, simultaneously. They are generally designed by creating an optical defect in the band gap of a structure with periodic modulation of the refractive index. Popular designs are based on a 2D PhC, where one or several holes are removed from the otherwise periodic lattice~\cite{Dharanipathy2014,Zhanga2016,Safavi-Naeini2010,Xiao2017}. The propagating mode of the waveguide at the center becomes evanescent at the edges and is thus confined, forming a resonant mode. Cavities are also obtained in 1D PhC and usually are made in a ridge waveguide in which a series of air-holes are etched. The two series are usually separated by a distance L forming the cavity, where light is trapped. 

In this article, we present the theoretical proposal of a CQED system based on thermal atoms coupled to 1D nano-beam cavity where a periodic spatial modulation in dielectric distribution is arranged over a dielectric nano-beam. The properties of such a photonic crystal have been exploited to provide sub-wavelength optical dispersion engineering. As suggested by our numerical calculations, this system is expected to show sufficiently strong coupling that even in the presence of the inhomogeneous broadening of the thermal atoms and the large transit time broadening, a strong coupling between the atom and the cavity photon is achievable. We also calculate some of the important atom-surface effects, like the Casimir-Polder potential, to incorporate the surface effect on the atom lines. Moreover, we extend the study further to include the behavior of more than one atom in the vicinity of the cavity and show that the temporal and spectral information can be used to distinguish between different cases.

\section{Photonic Crystal Cavity}
\begin{figure*}
\centering
\includegraphics[scale=1]{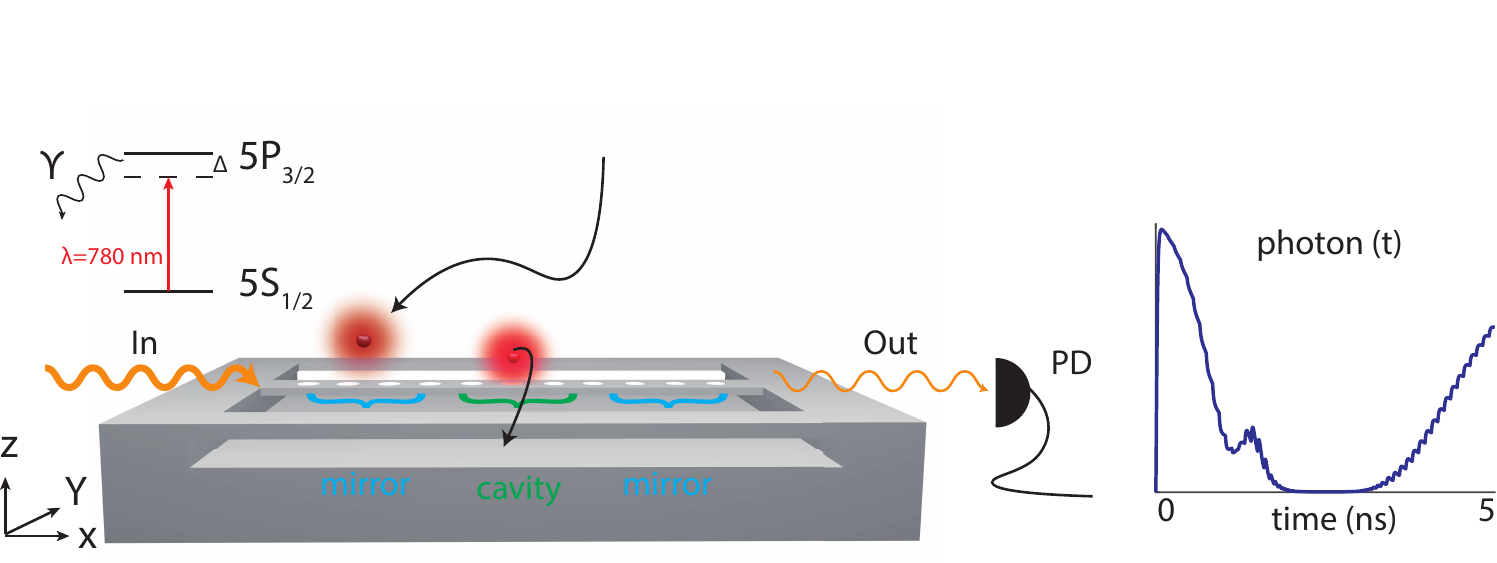}
\caption{\label{Fig1} (a) Schematics of the nano-beam photonic crystal cavity and the electronic energy levels of rubidium atoms flying along random trajectories in the vicinity of the structure. A pair of grating couplers at input and output will be used to excite the device with a coherent light (In) and collect the scattered light (Out) from the device and send it to a single-photon detector. The graph shows an example of the time-trace of the photons measured by photo diode.}
\end{figure*}

The structure of our interest is a suspended Silicon Nitrite (SiN) nano-beam whose refractive index has been modulated with a 1D array of air holes as schematically shown in Fig.~\ref{Fig1}. Rubidium (Rb) atoms with thermal velocity $\vec{v}$ and Maxwell-Boltzmann distribution fly in the vicinity of the nano-device and along random trajectories as denoted by black arrows in Fig.~\ref{Fig1}. The structure is assumed to be excited with coherent light via input grating couplers (In). The transmitted light after interacting with the atoms will be collected from another grating coupler at the output (Out) and sent to a fast single photo diode.  

\begin{figure*}
\centering
\includegraphics[scale=1]{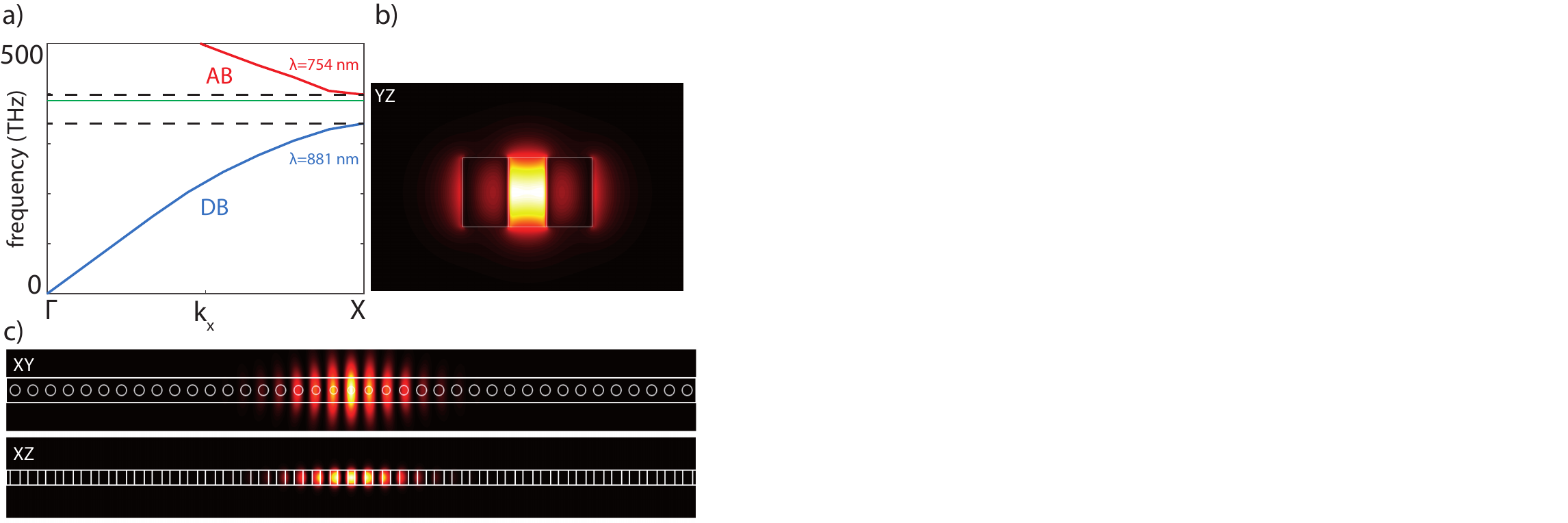}
\caption{\label{Fig2} (a) Photonic band-diagram of a 1D periodic hole array in a SiN nano-beam within the first Brillouin zone. The dashed lines show the photonic bandgap edges and DB and AB solid lines show the dielectric band and air band, respectively. The green line (horizontal solid line) shows the cavity resonance frequency with respect to the band edges. Electric field intensity profile of the cavity mode in (b) YX and (c) XY, XZ cross sections. The solid white lines in panels (b) , (c) show the physical boundaries of the structure.}
\end{figure*}

The periodic modulation of the refractive index of the nano-beam in 1D leads to a partial photonic bandgap along the x-direction. Within the frequencies of our interest, i.e. near infrared,  SiN can be treated as a dispersion-less material with fixed refractive index $n_{SiN}$ = 2.05~\cite{Tao2015}. Figure~\ref{Fig2}(a) shows the band structure of such 1D periodic structure, with $w$ = 420 nm, $h$ = 250 nm, $a$ = 325 nm, and $r$ = 91 nm, in the first Brillouin zone $[0 , ~\pi/a]$. With these parameters the structure supports a wide photonic bandgap with band edges at $\lambda$ = 881 nm and $\lambda$ = 754 nm for the dielectric (DB) and air band (AB), respectively. This wide bandgap implies that the periodic array serves as a good mirror for the photons within that energy range. These regions, indicated as mirror in Fig.~\ref{Fig1}, trap the photons for a long time in the cavity section ($\tau_{ph}$ is on the order of ns). By optimizing the positions and radii of the holes in the middle part a defect center can be realized that supports a single, high quality factor resonance. Since the atoms fly through the holes, the field intensity should be confined mostly inside the air holes, and hence the resonance frequency should be close to the air band (AB). In addition, to achieve the longest photon lifetime within the cavity, the spatial profile of the resonant mode should have a Gaussian distribution to minimize the leakage of the photons along $y,z$-directions (directions that light confinement is solely limited via total internal reflection). These conditions result in the radii of $r$ = 63, 67, 73, and 81 nm for the holes in the cavity region. These parameters lead to a resonant mode at 780 nm, i.e. $D_2$-line transition of Rb.
Figure~\ref{Fig2}(b),(c) show the field intensity profile of the cavity mode at three different cross sections. As can be seen the resonant mode is tightly confined within the cavity region and close to the nano-beam with a Gaussian profile along all directions. 

From these mode profiles we can calculate the relevant CQED parameters of this atom-cavity system. Our numerical calculations indicate the quaity factor of $Q$ = 65,000 equivalent to a photon decay rate of $\kappa_{ph} \approx$  4 $\times$ 10 $^{10}$ 1/s.

Aside from the Q-factor, mode volume is another important indication of the atom-photon coupling strength. For a dispersion-less materials like SiN the mode volume is determined by

\begin{equation}
V_{mode} = \frac{\int_V dV~ \epsilon(\vec{r}) |\vec{E}(\vec{r})|^2}{\textrm{Max}(\epsilon(\vec{r}) |\vec{E}(\vec{r})|^2)}
\end{equation}

where $\epsilon(\vec{r})$ is distribution of the structure permittivity as a function of position and $\vec{E}(\vec{r})$ is the electric field profile~\cite{Srinivasan2006}. 

Due to the sub-wavelength features of the mode profile the vacuum Rabi frequency is highly position-dependent and its maximum in the middle of the central hole is given by

\begin{equation}
g_{max} = \sqrt{\frac{\omega_{res}}{2\hbar \epsilon_0 V_{mode}}} \cdot d_{D_2}
\end{equation}

where in the above relation $\epsilon_0$ is the vacuum permittivity and $d_{D_2}$ = 3.584 $\times$ 10$^{-29}$ C$\cdot$m is the transition dipole moment of Rb at 780 nm~\cite{Steck2001}. 

The atom is flying through the device and will experience a transit-time broadening due to the finite time it has to interact with the field. For an atom at room temperature with the r.m.s velocity of $\sigma = \sqrt{3k_B T/m_{Rb}}\approx$ = 300 m/s the broadening due to the finite interaction time with the cavity field can be approximated as 

\begin{equation}
\tau_{int} = \frac{h}{\sigma} \approx 1 ~ ns
\end{equation} 

where in the above equation $h$ is the slab height which is the shortest interaction length of the atom and the cavity. This is just an order of magnitude calculation to estimate the achievable atom-photon cooperativity expected from this cavity. The effect of this transient interaction between the atom and the localized field of the cavity is exactly incorporated in the Monte-Carlo calculations presented in section III. 

From these values we can estimate the cooperativity of this atom-cavity system, a parameter that indicates the ratio between all the coherent and incoherent effects in a coupled system and is determined as

\begin{equation}
C  = \frac{g_{max}}{\sqrt{\kappa_{ph} \Gamma_{atom}}}
\end{equation}

Table~\ref{Table1} summarizes the important parameters of this cavity-atom system.

\subsection{Green's function and LDOS}
As was stated in the previous section the cavity has a large quality factor (Q) and a small mode volume hence the radiation properties of an atom in the vicinity of the structure would be strongly modified. One of the interesting aspects of nano-photonic devices is their ability to substantially modify the local density of the optical states (LDOS) due to the large gradient of the field profile on the order of or below the resonance wavelength. 

The modification of an emitter in the vicinity of the structure occurs in two ways: 1- via ng the decay rate of the emitter by a factor known as the Purcell factor. 2- via modifying the electronic levels of the emitter. Since the refractive index of the device is rather large there would be a noticeable dipole-dipole interaction potential between the emitter and its image. This Casimir-Polder interaction induces a shift in the atomic line~\cite{Scheel2015}. 
As both of these phenomena are related to the Green's function of the device, in this section we present the Green's function and calculate the corresponding Casimir-Polder potential out of it.

The electromagnetic Green's tensor denoted as $\textbf{G}(r,r';\omega)$, is the electric field at location $r$ generated by an infinitesimal current moment at location $r'$ and at frequency $\omega$. In  other words, it is the impulse response of the Helmholtz operator determined via the following equation:

\begin{equation}
\left(\nabla^2 - (\frac{\omega}{c})^2 \epsilon(r) \right) \textbf{G}(r,r';\omega)= \vec{\delta}(r-r')
\end{equation}

where in the above equation $\epsilon(r)$ is the spatial distribution of the permittivity set by the nano-photonic device geometry. 

Aside from a few cases with specific symmetries, in general the Green's tensor of most systems should be determined numerically. In this study we employed $Lumerical$ an FDTD-based commercial software to solve this equation numerically. 

The Green's tensor is also an indication of the radiated power from the dipole in the system. Therefore, the larger the coupling to the modes the larger radiated power by the dipole. In this nano-beam geometry the cavity mode is mainly polarized along the y-direction, specifically in the $XY$ plane of symmetry. This implies that the best coupling to various dipole orientations can be obtained for a $y$-directed dipole, the dipole oriented along the beam width. For the other two orientations the radiated power is substantially lower as the dipole and the cavity modes do not couple efficiently. This qualitative prediction can be clearly observed in the numerical calculations presented in Fig.~\ref{Fig3}(a) where the calculated Purcell factor for all of the $x,y$ and $z$-oriented dipoles has been shown as a function of the dipole wavelength. As can be observed the power emitted from a $y$-directed dipole substantially increases when the dipole energy approaches the cavity resonance wavelength. Moreover, the emitted power is noticeably higher compared to the power emitted from $x,z$-directed dipoles. Note that the radiated powers from the latter dipoles are not strictly zero since aside from the cavity mode there are continuum of leaky modes that the dipoles could always couple to. 
The finite spikes in the radiation spectrum of the $x,z$-oriented electric dipoles are due to some parasitic cavity effects in a finite-sized structure we considered in all of the simulations.  

\subsection{Casimir-Polder potential}
When an emitter radiates in the vicinity of a large-index dielectric surface its radiation properties not only will be affected by the change of the local density of the optical states (LDOS) as discussed in the previous section, but also its electronic levels will be affected by a dipole-dipole interaction potential between the atoms and its induced dipole image. This phenomena known as Casimir-Polder potential is closely related to the modification of the emission properties of the dipole locally and is related to the Green's function via the following relation~\cite{Buhmann2012}

\begin{equation}
U_{CP}(r) = - \frac{\hbar}{2\pi} ~\mu_0 \int_0^\infty \textit{Im}\left(\textit{Tr}~ (\alpha(\omega) \textbf{G}_{sc}(r,r;\omega)) ~\omega^2 \right) ~ d\omega
\end{equation}

where $\textbf{G}_{sc}(r,r;\omega)$ is the scattered Green's tensor as defined in the previous section and $\alpha(\omega)$ is the dynamical polarizability tensor of the emitter at frequency $\omega$. 

\begin{figure*}
\includegraphics[scale=0.8]{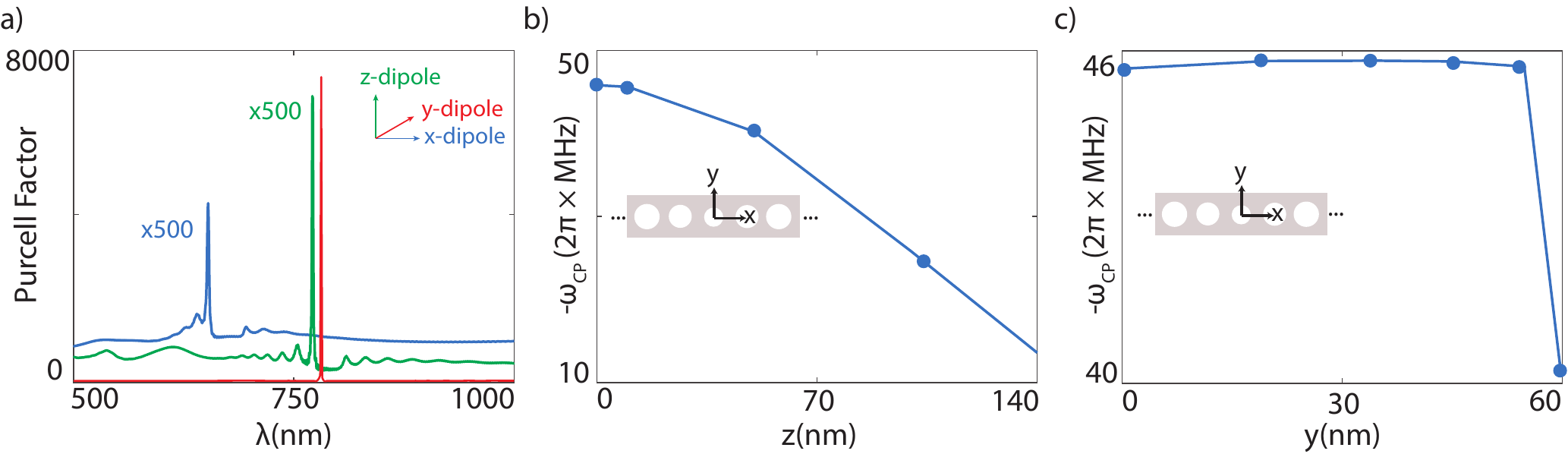}
\caption{\label{Fig3} (a) Purcell factor enhancement of a radiating electric dipole at the center of the cavity, i.e. $x=y=z$ = 0 as a function of the dipole wavelength. The blue, red, and green lines correspond to the $x,y$, and $z$-oriented electric dipoles. Purcell factors of $x$- and $z$-oriented dipoles are multiplied by a factor 500 to have comparable values to the $y$-dipole case.(b) Casimir-Polder induced line shift as a function of z, away from the device surface, for an electric dipole at $x=y$ = 0. The slab height $h$ = 250 nm. The induced potential is calculated for a dipole emitting at D$_2$ line of rubidium. (c)  Casimir-Polder induced line shift as a function of y, along the beam width, for an electric dipole at $x=z$ = 0. The radius of the central hole $r$ = 63 nm. The filled circles in each case show the simulated data points.}
\end{figure*}

In general, the induced force is related to all of the real and virtual transitions of the atoms. However, as suggested by the results in Fig.~\ref{Fig3}(a), the Green's function of this cavity is mainly peaked at the resonance wavelength (i.e. $\lambda_{res}$) and has substantially lower values away from it. For such a high-Q cavity we can limit ourselves to the behavior of the Green's function in the vicinity of the resonance. Moreover, since this cavity is single mode, the Green's function does not have any further features away from this resonance and the approximation is valid through the whole frequency range. Therefore, the Green's function of the cavity can be properly approximated with a single Lorentzian within the whole frequency range. 

Employing this approximation and calculating the polarizability from the transition dipole moments and transition frequencies, we have calculated the Casimir-Polder potential at several points close to the device. In Fig.~\ref{Fig3}(b),(c) we present some of the calculated results away from the structure along the $z$-direction, and away from the cavity towards the hole edge along the $y$-direction, respectively. As can be seen, due to the induced dipole image a Rb atom radiating at $D_2$ experiences about tens of MHz line shifts in the transition frequency. Also due to the attractive force between the dipole and its image the shift in the energy is always negative. Comparing these values with the transit broadening an atom experiences as it flies though the structure it can be concluded that the Casimir-Polder potential does not have a large effect. Indeed, the effect becomes more and more pronounced for the slower atoms where the energy shift can lead to measurable detuning between the atom and cavity hence, diminishing the strong coupling between the atom and cavity photon.

\section{Atom-Light Interactions and Monte-Carlo Simulations}
In previous sections we conducted some orders of magnitude estimation of the atom photon coupling as well as more quantitative investigations on the Casimir-Polder potential to explore the behavior of a moving atom in the vicinity of well-designed high-Q PCC with small mode volume. Those analyses suggest that the achievable coupling between a flying atom and a photon cavity is strong enough to expect CQED behavior from this hybrid system even though the moving atom typically experiences large transit time broadening. To fully evaluate the performance of the hybrid system however, one needs to treat the atom-photon coupling more accurately. In this section we develop a Monte Carlo scheme that combines the Casimir-Polder potential derived in the previous with full quantum mechanical description of the atom-photon coupling to investigate the interaction of a moving atom with the nano-cavity. 
For that, we investigate the motion of a thermal atom with a random velocity $\vec{v}$ following a random trajectory $\vec{r}(t)$ determined by its velocity. The evolution of the joint density matrix of the atom and photon at each time is determined via the following Liouville's equation 

\begin{widetext}
\begin{equation*}
\frac{d\rho}{dt} = -i\left[H , \rho\right] + \kappa \left(a \rho a^\dagger - \frac{1}{2}\left\lbrace a^\dagger a , \rho \right\rbrace  \right) 
+ \Gamma \left(\sigma^- \rho \sigma^+ - \frac{1}{2}\left\lbrace \sigma^+ \sigma^- , \rho \right\rbrace  \right)
\end{equation*}
\end{widetext}

where ($\kappa, \Gamma$) are the photon decay rate from the cavity and the atom lifetime in the excited state, respectively. $a,a^\dagger$ are the annihilation and creation operators of the photon field, and $\sigma^- , \sigma^+$ are the atomic lowering and raising operators, respectively.

Atom-photon interaction can be best studied with Fock states. However, since generating the Fock states is a hard task we investigate the scenario when photons are injected with a weak, coherent excitation at rate $\epsilon_p$. Therefore, the total Hamitonian in the rotated frame of the pump, is given via the following driven Jaynes-Cumming form as

\begin{widetext}
\begin{equation*}
H = \left(\omega_c - \omega_p \right) a^\dagger a +  \left(\omega_a(t) - \omega_p \right) \sigma^+ \sigma^- + g(t)  \left(a \sigma^+ + \sigma^- a^\dagger \right) + \epsilon_p \left( a + a^\dagger \right)
\end{equation*}
\end{widetext}

where $\omega_c,\omega_a, \omega_p$ are the cavity, atom, and laser frequency respectively and $g(t)$ is the vacuum Rabi coupling of the atom-cavity which depends on the atom position and varies as the atom moves along the device. Also, due to the Casimir-Polder effect, the detuning between the atom and the laser is position and therefore time dependent as well. The time-dependent state of the atom and cavity photons describe a time-varying entanglement between the atom and cavity systems.

\begin{figure*}
\includegraphics[scale=0.8]{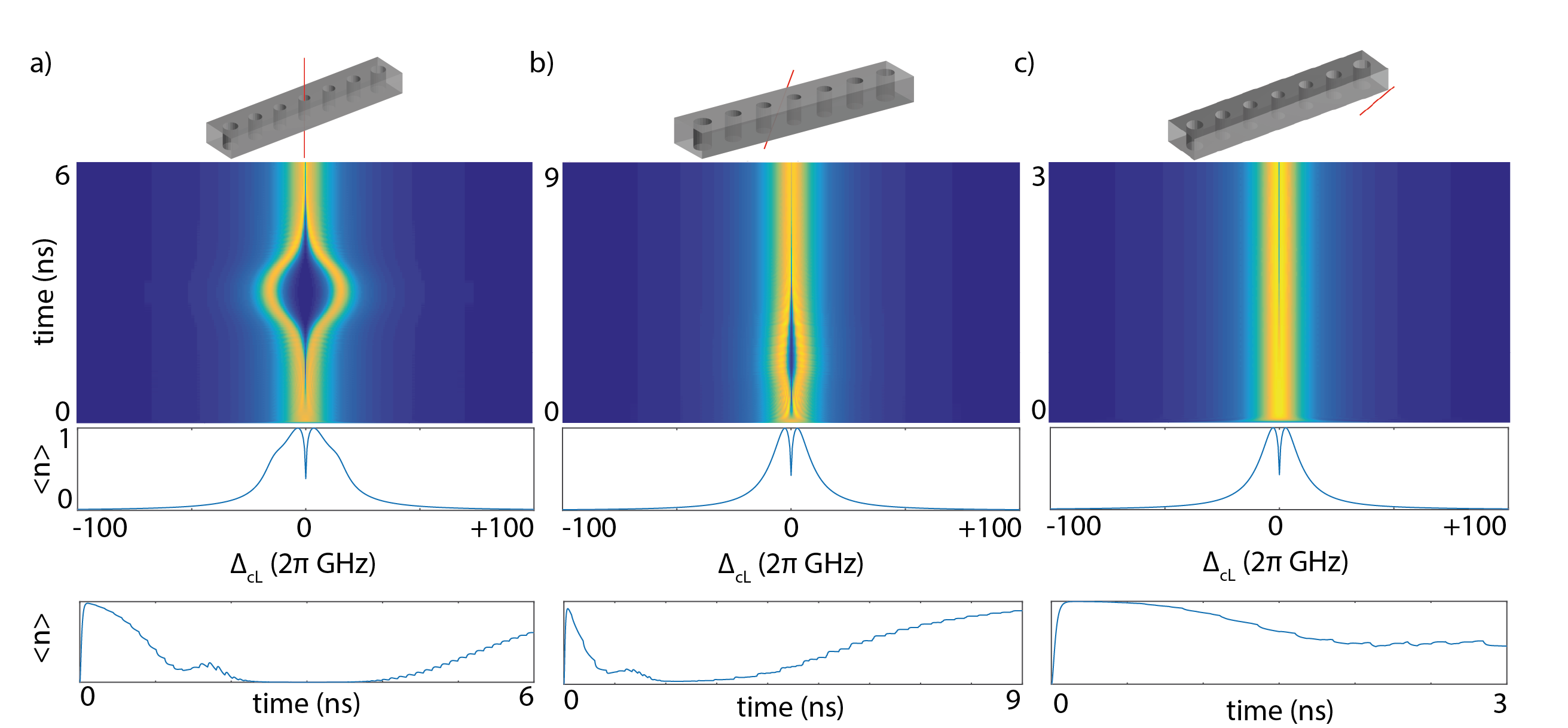}
\caption{\label{Fig4} Instantaneous photon numbers $\braket{a^\dagger a}$ of moving atoms as a function of cavity-laser detuning ($\Delta_{cL}$) for an atom (a) $r_0$ = (0 0 600) nm and $v_0$ = (0 0 -200) m/s , (b) $r_0$ = (100 150 300) nm and $v_0$ = (0 50 -100) m/s, and (c) $r_0$ = (800 -200 -500) nm and $v_0$ = (120 -40 100) m/s. In each case the simulation has been terminated when an atom crashes onto the device or the simulation boundaries. The schematics on top in each case show the atom trajectory alongside the photonic crystal cavity. In each sub-figure the color-map panel shows the real-time variation of the intra-cavity photon number, the middle row shows the averaged signal over time as a function of the cavity-laser detuning, and lower panel shows the time signal at zero detuning ($\Delta_{cL}$ = 0).}
\end{figure*}

Figure~\ref{Fig4} shows the results of the Monte Carlo simulation for a moving atom in the vicinity of PCC for three different velocities and initial position. The joint density matrix has been evolved according to the above equation and the instantaneous expectation value of the intra-cavity photon numbers $\braket{a^\dagger a}$ has been calculated for different cavity-laser detuning values ($\Delta_{cL}$) while the atom moves in the vicinity of the cavity. 
The interaction time and the coupling strength between atom and cavity strongly depend on the atom velocity and its initial position, hence the behavior drastically changes along different trajectories. The time-averaged photon number as a function of cavity-laser detuning has been shown in the second row for each case. The third row manifests the change of the intra-cavity photon number as a function of time when $\Delta_{cL}$=0. As can be observed the number of photons in the cavity drops when the coupling between the atom and the cavity photon is strong enough to induce a large detuning between the input laser frequency and the frequencies of the dressed states. In other words, the cavity becomes opaque to the input laser for a finite amount of time. The inverse scenario is observable as well. At large cavity-laser detuning the presence of atom makes the cavity transparent for a while.

\begin{figure*}
\includegraphics[scale=0.8]{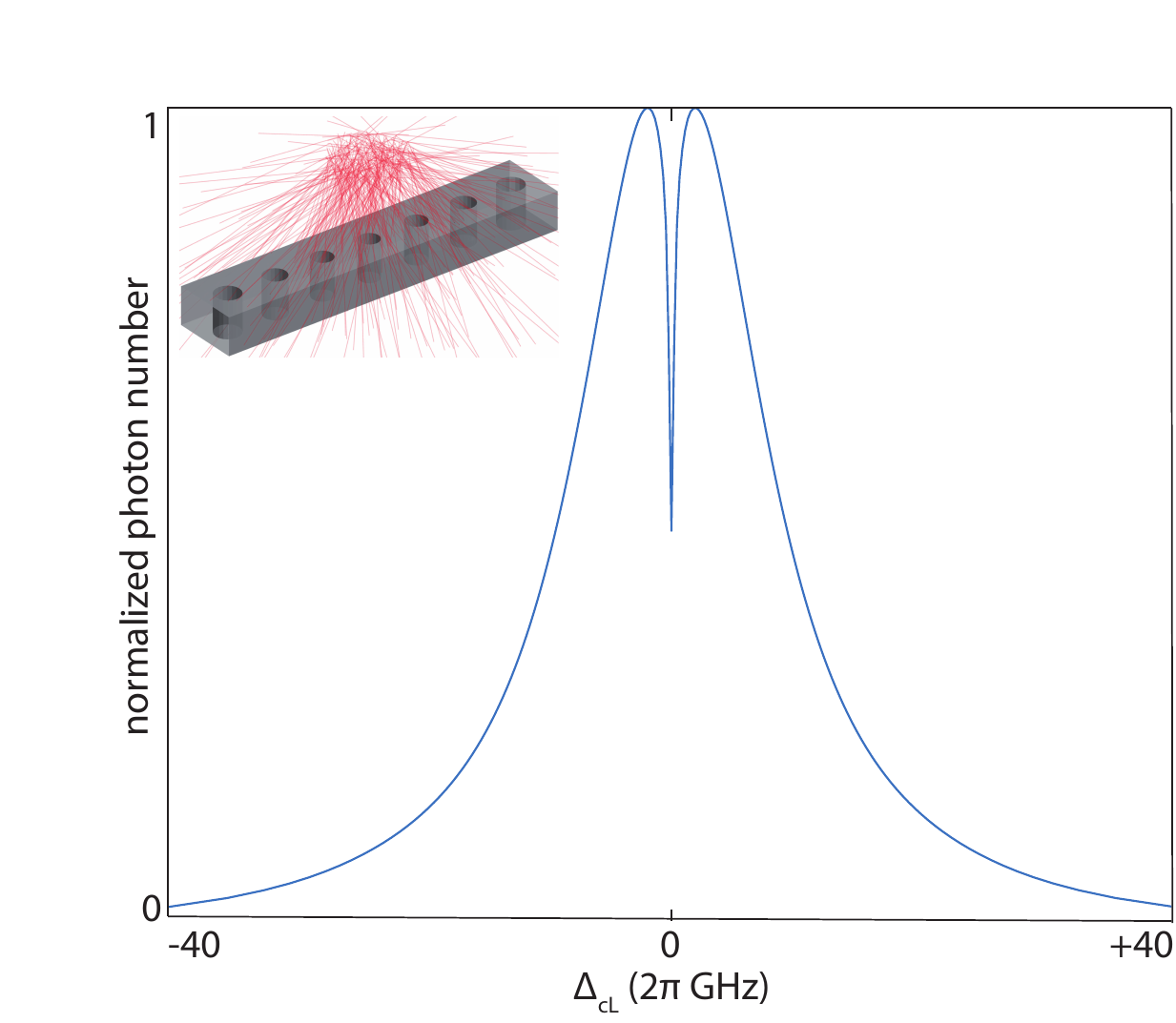}
\caption{\label{Fig5} Normalized photon number as a function of laser-cavity detuning $\Delta_{cL}$, averaged over 100,000 trajectories generated from light induced atomic desorption (LIAD) process. The inset shows some of the trajectories of released atoms.}
\end{figure*}

To produce a gated cloud of atoms one can employ the light-induced atomic-desorption (LIAD) process. In the LIAD process, also known as the photo-electric effect for atoms, high energy laser pulses hit the wall of an atomic vapor cell, which is coated with alkali atoms and desorb them from the surface~\cite{Meucci1994,Klempt2006,Campo2015,Barker2018}. The inset of Fig.~\ref{Fig5} shows the typical trajectory of atoms released from the wall via LIAD process. Figure~\ref{Fig5} shows the normalized intra-cavity photon number as a function of cavity-laser detuning after averaging over 100,000 atom trajectories.
Due to the random paths, which give different atom-photon coupling, some of the strong coupling signatures have been smeared out. However, one can still observe a clear decrease of intra-cavity photon number at $\Delta_{cL}$ = 0 which is a hallmark of strong coupling between the cavity photons and the atom ensemble.

\begin{figure*}
\includegraphics[scale=0.8]{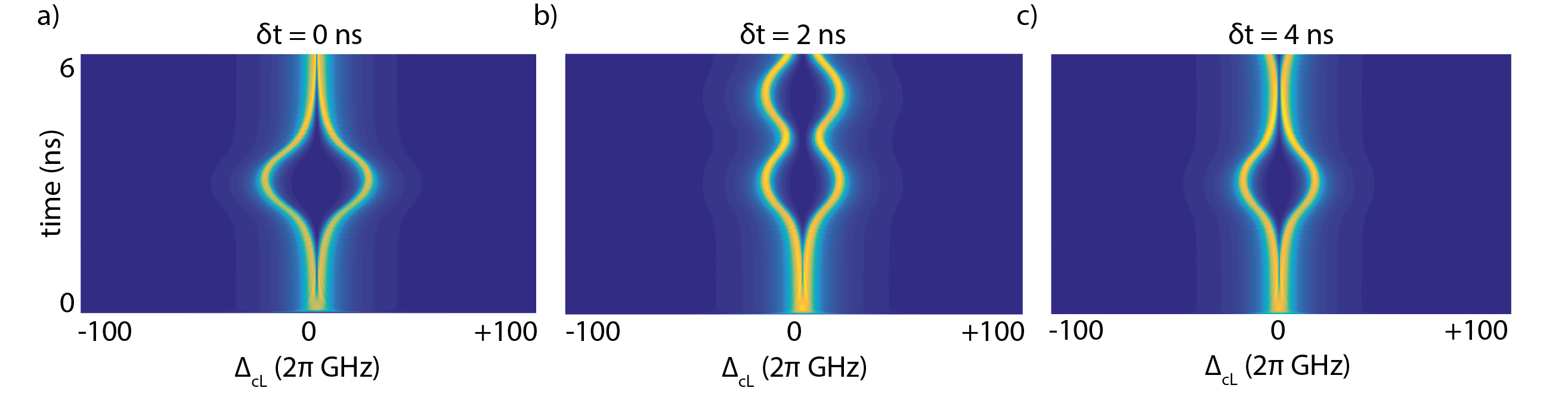}
\caption{\label{Fig6} Instantaneous intra-cavity photon number as function of cavity-laser detuning ($\Delta_{cL}$) when two atoms fly through the central hole. Both atoms have the same velocity of $\vec{v} = -200\hat{z}$ m/s and move towards the cavity along the same trajectory. However, they are delayed by $\delta t$ (a) 0 ns, (b) 2 ns and (c) 4 ns.}
\end{figure*}

The analysis can be extended further to investigate the behavior of multiple atoms interacting with the same cavity mode. Figure~\ref{Fig6}(a)-(c) shows the instantaneous number of intra-cavity photons as a function of cavity-laser detuning for three different delays between two atoms for $\delta t = 0, 2, 4$ ns, respectively. The complete study of the many-atom cavity coupling in the dense atomic media is the topic of follow up studies and as very different scenarios are imaginable here we only limit ourselves to atoms flying through the central hole, where the interaction is the strongest. In all of these cases the atoms follow the same trajectory along the line passing through the center of the cavity with the velocity of $\vec{v} = -200 \hat{z} ~ m/s$. As can be seen depending on the delay between atoms the behavior is substantially different. When there is no delay and both atoms fly in the vicinity of the cavity simultaneously the behavior is similar to the single atom case (Fig~\ref{Fig4}(a)) except an increase in the effective Rabi coupling by $\sqrt{2}$. When there is a finite delay between atoms, as depicted in Fig.~\ref{Fig6}(b),(c), the first atom can make the cavity opaque while the second atom can allow the photons to enter the cavity again. This behavior, which is strongly delay-dependent is depicted in panel (b) and (c) where multiple oscillations in the instantaneous photon number can be observed. For longer delays the first atom is already far away from the cavity and its effect is so negligible that the second atom can be treated almost, independently.

\section{Conclusion}
In this work we proposed a new hybrid system for CQED studies with thermal atoms. To overcome the typical problem of short interaction time between flying atoms and the cavity mode we designed a high quality factor photonic crystal cavity that provides large vacuum Rabi coupling and cooperativity. We studied the effect of cavity and LDOS modification in controlling the decay rate. In addition, we calculated the surface effect and Casimir-Polder potential in changing the atomic lines. Using a Monte Carlo algorithm combining the full quantum-mechanical description of atom-photon interaction with Casimir-Polder effects, we investigated the feasibility of observing coherent coupling between a flying atom and the cavity photon. Our results predict that the attainable coupling can be large enough to achieve the strong coupling regime in spite of all decoherence and transit time effects. In addition, we extended the study to more than one atom case and investigated multiple atom-induced transparency occasions that can be achieved in this case.

Our designs and analysis set the foundations for investigating atom-photon interactions in more complicated nano-photonic devices that support specific features such as topology or chirality~\cite{Khanikaev2017}. Furthermore, due the large attainable LDOS in nano-devices the coupling of atoms to non-desired modes will be substantially suppressed and photon mediated coupling between atoms are expected to be enhanced. Interfacing atoms with custom nano-photonic devices is a burgeoning emerging field and new eras of quantum optics are ahead, which leads to novel types of interaction between the atoms and the engineered photonic states.

\section*{Acknowledgment}
The authors thank the fruitful discussions with Stefan Scheel and Helge Dobbertin. H.A. acknowledges the financial support from IQST and Eliteprogramm of Baden-W\"urttemberg Stiftung.

\bibliographystyle{ieeetr}
\bibliography{ref}

\newpage

\begin{table}
\begin{center}
\begin{tabular}{|c|c|c|c|c|c|}
\hline
$\lambda_{res}$ & $V_{mode}$ & Q & $g_{max}$ & $\kappa$ & C \\
\hline
780 nm & 0.08~$\lambda_{res}^3$ & 65,000 & 2$\pi \times$ 15 GHz &  4$\times 10^{10}$ s$^{-1}$ & 18 \\
\hline
\end{tabular}
\end{center}
\caption{\label{Table1} CQED parameters of the nano-beam photonic crystal cavity.}
\end{table}

\end{document}